\documentclass{moriond}
\usepackage{graphicx,subfigure}

\def\be{\begin{equation}}
\def\ee{\end{equation}}
\def\bea{\begin{eqnarray}}
\def\eea{\end{eqnarray}}

\usepackage{amsmath}
\usepackage[square]{natbib}
\setcitestyle{numbers}

\newcommand{\Photo}{}

\begin{document}
\vspace*{0.1cm}
\title{First Dark Matter Search Results from the Large Underground Xenon (LUX) Experiment}

\author{ C.H.~Faham, for the LUX Collaboration}

\address{Lawrence Berkeley National Laboratory, 1 Cyclotron Rd., Berkeley CA 94720, USA}

\maketitle\abstracts{The Large Underground Xenon (LUX) dark matter experiment is operating 1.5~km underground at the Sanford Underground Research Facility in Lead, South Dakota, USA. In 2013, the experiment had a WIMP search exposure of 10,091~kg-days over a period of 85.3~live days. This first dark matter search placed the world's most stringent limits on WIMP-nucleon interaction cross-sections over a wide range of WIMP masses, and is in tension with signal hints of low-mass WIMPs from DAMA, CoGeNT and CDMS-II Si. LUX will commence a 300~day run in 2014 that will improve the sensitivity by a factor of 5. Low-energy calibrations obtained from a neutron double-scattering technique will further constrain and reduce systematics, particularly for low WIMP masses.}

\section{Introduction}

The existence of non-baryonic cold dark matter is well established through a wealth of complementary observations such as Supernova Type Ia, the Cosmic Microwave Background, Big Bang Nucleosynthesis, Baryon Acoustic Oscillations and the kinematics of galaxies and galaxy clusters \cite{Beringer:1900zz}. Despite the overwhelming cosmological evidence, the particle nature of dark matter remains unknown. One of the leading candidates for dark matter is the Weakly Interacting Massive Particle (WIMP), which is expected to interact with a terrestrial detector through low-energy ($\sim$keV scale) nuclear recoils (NR). Direct-detection experiments aim to identify these faint and rare recoils by utilizing a large target mass, achieving a low energy threshold and reducing background radioactivity levels, mainly from electron recoils (ER) and neutrons. Furthermore, direct-detection experiments need to operate deep underground in order to significantly reduce the otherwise insurmountable cosmic-ray background rate.

\section{The LUX Detector}

The Large Underground Xenon (LUX) dark matter experiment has been operating 1.5~km (4300~mwe) underground at the Sanford Underground Research Facility (SURF) in Lead, South Dakota, USA since 2012 \cite{Akerib:2013tjd}. LUX is a two-phase xenon time-projection chamber (TPC) constructed from ultra-low radioactivity materials. The detector consists of 370~kg of liquid xenon, 250~kg of which comprises the 47~cm diameter by 48~cm high cylindrical active volume. The detector is immersed in a 7.6~m diameter by 6.1~m high water tank to reduce the background from ambient radioactivity.

A particle interaction in the liquid xenon volume, through either an ER or an NR, produces primary scintillation photons (S1 signal) and ionization electrons, which are drifted with an electric field into the gas phase, where they produce secondary scintillation light (S2) via electroluminescence. A total of 122 Hamamatsu R8778 low-radioactivity photomultiplier tubes (PMTs) are used for the detection of both S1 and S2 signals. The PMTs are split into top and bottom arrays of 61 PMTs each. The LUX PMTs have an average photon detection efficiency\footnote{The PMT photon detection efficiency, or DE, is the product of the often quoted quantum efficiency (QE) times the first dynode electron collection efficiency (CE).} of about 30\% at the xenon scintillation wavelength of 175~nm. The inner walls of the detector and the space between PMTs are lined with polytetrafluoroethylene (PTFE), which has been measured in LUX to be $>$95\% reflective at 175~nm when immersed in liquid xenon. These detection factors, combined with the reflectivity of the grids' stainless steel wires and a finite photon absorption length in xenon, yield a measured 14\% LUX photon detection efficiency for events at the center of the detector. This large photon detection efficiency gives LUX good sensitivity to WIMP masses above 6~GeV/c$^2$. For further details about the LUX experiment, refer to \cite{Akerib:2012ys}.

\begin{figure}
\begin{center}
\includegraphics[width=0.85\textwidth]{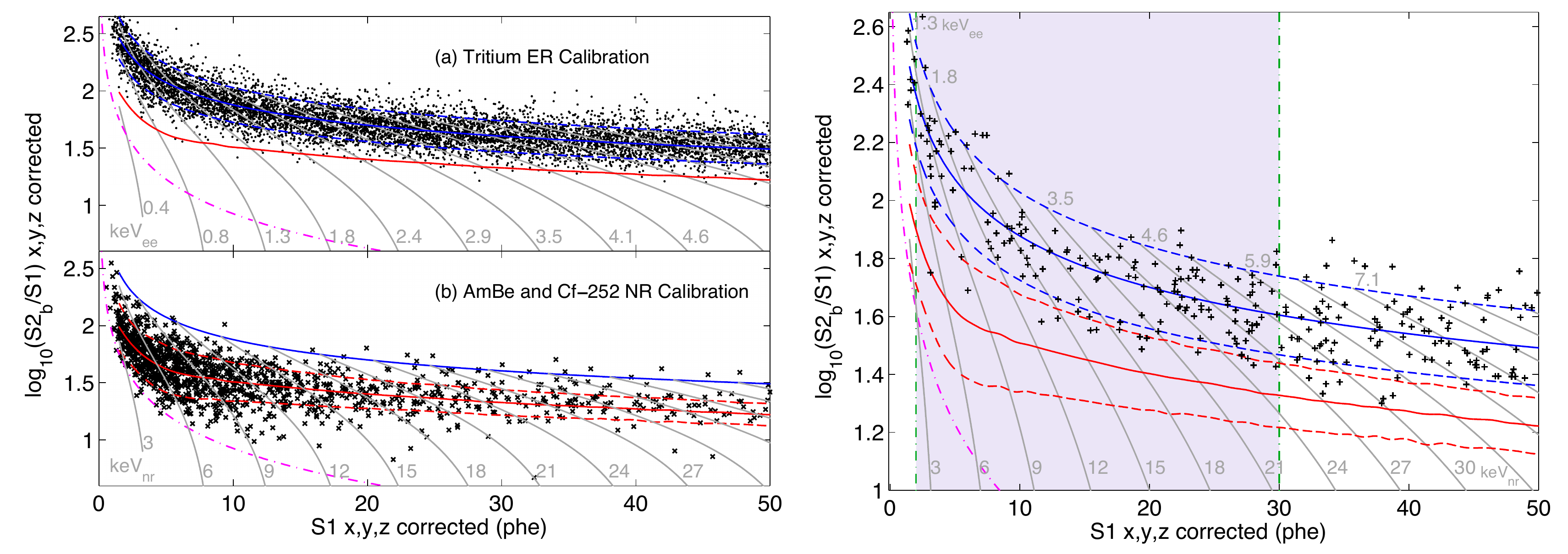}
\caption{\textit{Left}: The LUX low-energy response to tritium ER (panel a) and AmBe and $^{252}$Cf NR (panel b) calibrations inside the fiducial volume. The ER band mean (solid blue) and $\pm1.28\sigma$ (dashed blue) of the parametrized tritium distribution are shown in panel b. The NR band mean (solid red) and $\pm1.28\sigma$ (dashed red) obtained from simulations are shown in panel b. The approximate location of the 200~phe S2 cut is shown with a dot-dashed magenta line in both panels. \textit{Right}: The fiducial volume events passing all cuts in the 2-30~phe S1 range (shaded purple) during the LUX 85.3 live-day WIMP search. The same parametrized ER and NR band distributions from the left panel are shown. A Profile Likelihood Ratio analysis shows that all events are consistent with the ER background-only hypothesis.}
\label{fig:LUXResults}
\end{center}
\end{figure}

\section{The First LUX Dark Matter Search and Limits on WIMP-Nucleon Interactions}

A total of 85.3 live-days of background WIMP search data were acquired between the end of April and the beginning of August 2013. During this run, the background rate inside the fiducial volume in the energy range of interest (2-30~phe S1 signals) was measured to be $3.6\pm0.3$~mDRU (mDRU = $10^{-3}$ counts/keV/kg/day), which is the lowest for any xenon TPC thus far. About half of the fiducial background rate are low-energy scatters from gamma rays (mostly originating from residual radioactivity in the PMTs). Other sources of background are residual $^{85}$Kr in the xenon (rendered subdominant by using chromatographic separation \cite{Akerib:2013tjd}), $^{214}$Pb naked $\beta^-$ from the $^{222}$Rn chain and x-rays from the cosmogenically produced $^{127}$Xe (t$_{1/2}$ = 36.4~days), which decayed throughout the WIMP search run.

The detector was extensively calibrated with internal sources (for ER) and external sources (for NR). The internal sources, $^{83\text{m}}$Kr and tritiated methane, were injected in gaseous form into the xenon circulation stream. These internal sources have the advantage of spreading evenly throughout the active volume, providing a homogeneous calibration. The mono-energetic 9.4~keV and 32.1~keV energy depositions of $^{83\text{m}}$Kr were used to constantly monitor the electron drift attenuation length, the light yield and the x,y,z detector response corrections. The novel tritiated methane $\beta^-$ source (endpoint of $\sim$18~keV) provided the ER response of the detector at low energies, as shown in the left panel (a) in Fig. \ref{fig:LUXResults} in log$_{10}$(S2$_\text{b}$/S1)~vs.~S1 space (the ``b" subscript denotes that only the bottom PMT signals were used in order to eliminate systematics from two deactivated top PMTs). The mean (blue solid) and $\pm1.28\sigma$ (blue dashed) contours used to characterize the ER band from this calibration are shown on top of the data. AmBe and $^{252}$Cf external neutron sources were used to calibrate the NR response of the detector, shown in the left panel (b) in Fig. \ref{fig:LUXResults}. The mean (red solid) and $\pm1.28\sigma$ (red dashed) NR band parametrization was derived from the NEST simulation model \cite{Szydagis:2013sih}.

\begin{figure}
\begin{center}
\subfigure{\includegraphics[width=0.45\textwidth]{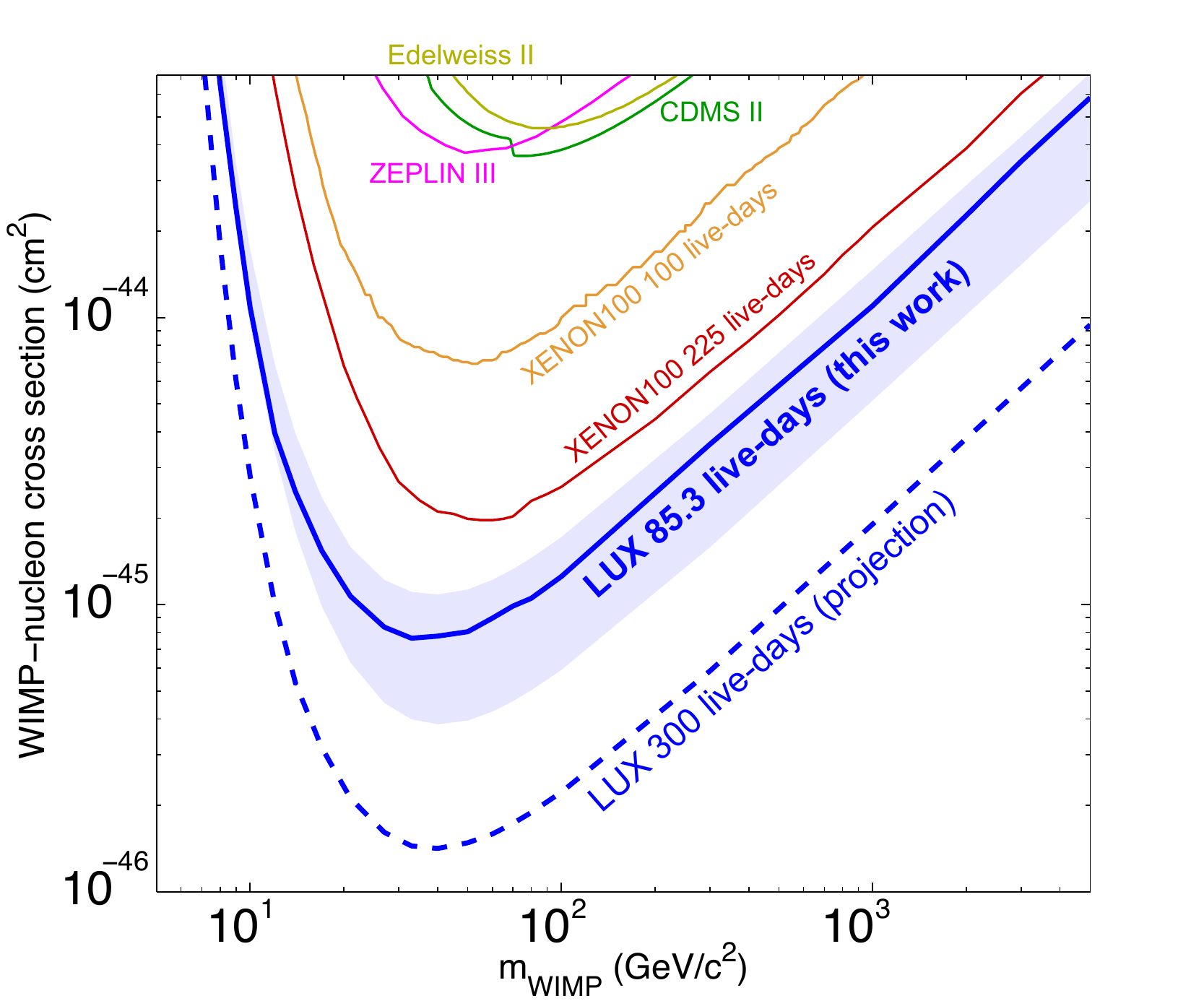}}
\subfigure{\includegraphics[width=0.45\textwidth]{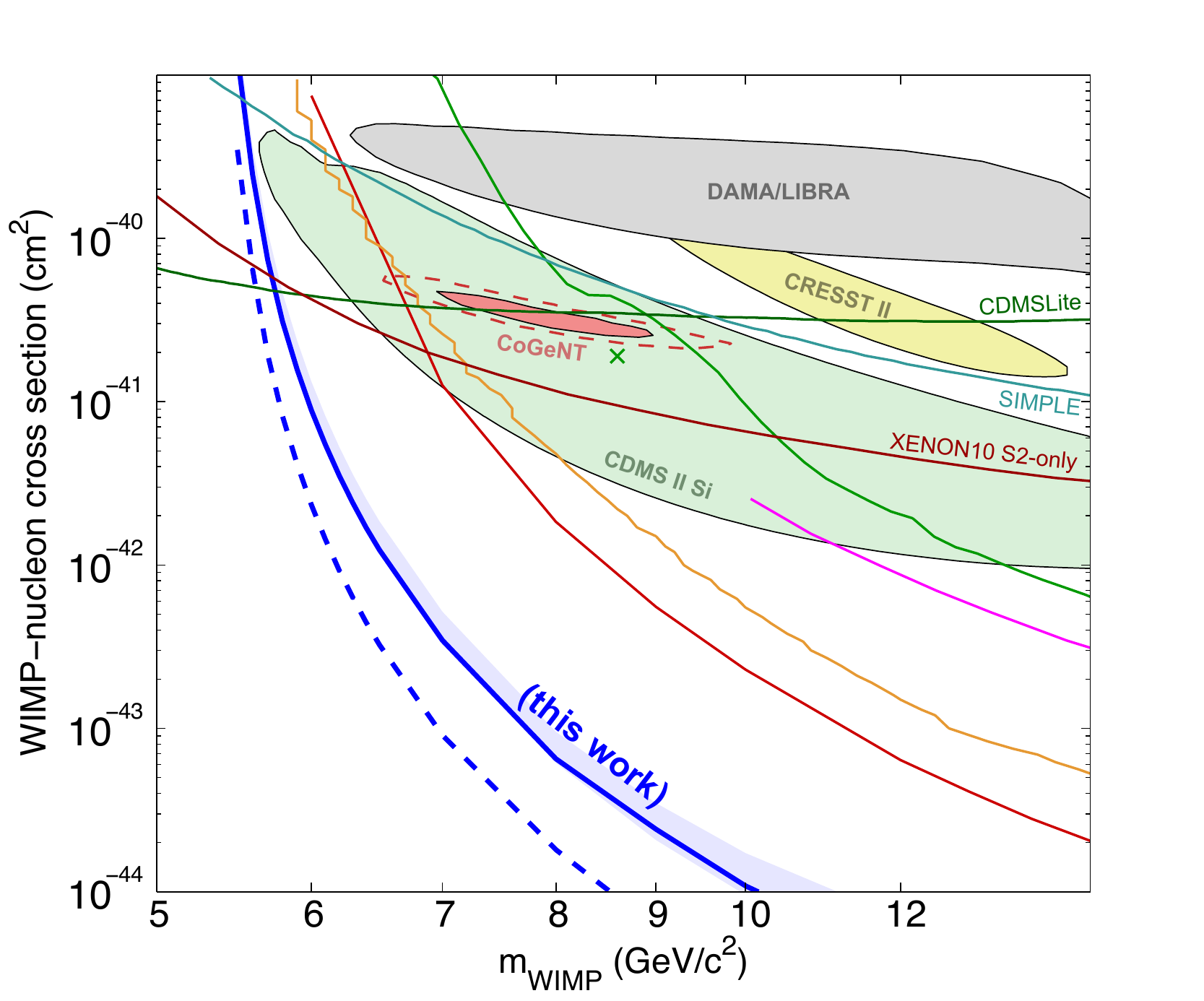}}
\caption{\textit{Left}: 90\% CL spin-independent WIMP exclusion limits shown the LUX 85.3 live-day result (solid blue) and the 300-day projection (dashed blue). \textit{Right}: Close-up view of exclusion plot in the low-mass regime showing the tension between the LUX result and previous hints of low-mass WIMP signals.}
\label{fig:LUXLimits}
\end{center}
\end{figure}

The WIMP search analysis cuts for this unblind analysis were kept minimal, with a focus on maintaining a high acceptance. Single-scatter interactions (one S1 and one S2) in the liquid xenon with areas between 2-30~phe for the x,y,z corrected S1 signal were selected, which approximately corresponds to 3-25~keV$_{\text{nr}}$ or about 0.9-5.3~keV${_\text{ee}}$, where the subscripts represent the energy scales for NR and ER, respectively.\footnote{For the same energy, a NR produces less signal than an ER due to the fact that the former has a large energy loss fraction in the form of heat, which produces no photons or electrons.} The upper bound of 30~phe was chosen to avoid contamination from the 5~keV x-ray from $^{127}$Xe. The fiducial volume was defined as the inner 18~cm in radius and a drift time between 38-305~$\mu$s (roughly 7-47~cm above the bottom PMT array). The fiducial mass enclosed by the aforementioned bounds was calculated to be $118.3\pm6.5$~kg from the tritium calibration. An analysis threshold of 200~phe ($\sim$8~extracted electrons) was used to exclude small S2 signals with poor x,y position reconstruction. The S2 finding efficiency at 200~phe is $>$99\%. The overall WIMP detection efficiencies after all cuts were roughly 17\% at 3~keV$_{nr}$, 50\% at 4.3~keV$_{nr}$ and $>95$\% above 7.5~keV$_{nr}$.

A total of 160 events passed the selection criteria, which are shown inside the purple shaded region in the right panel of Fig. \ref{fig:LUXResults}. A Profile Likelihood Ratio (PLR) analysis utilized the distribution of measured background and expected signal as a function of radius, depth, S1 and S2 parameter spaces in order to attempt to reject the null (background-only) hypothesis. For further details about the PLR limit, see \cite{Akerib:2013tjd} and \cite{Szydagis:2014xog}. The PLR result could not reject this null hypothesis with a p-value of 0.35, and 90\% confidence spin-independent WIMP exclusion limits were placed as a function of WIMP-nucleon cross-section and WIMP mass as shown in Fig. \ref{fig:LUXLimits}. The WIMP exclusion limits set by LUX provide a significant improvement in sensitivity over existing limits. In particular, the LUX low-mass WIMP sensitivity shown in the right panel of Fig. \ref{fig:LUXLimits} improves on the previous best limit set by XENON100 by more than a factor of 20 above 6~GeV/c$^2$. These low-mass limits do not support the near-threshold signal hints seen by DAMA \cite{Bernabei:2008yi}, CoGeNT \cite{Aalseth:2012if} and CDMS-II Si \cite{Agnese:2013rvf}. 

The WIMP exclusion limit in LUX was derived using a conservative xenon response to NR at low energies, which placed an unphysical cutoff in the signal yields for electrons and photons below 3~keV$_\text{nr}$, the lowest calibration point available at the time of the limit calculation. New measurements from a DD neutron generator show available signal below this imposed cutoff (measured down to 0.7~keV$_\text{nr}$ for the ionization channel) \cite{Verbus:2014dm}.

\section{Conclusions and Outlook}

During an 85.3~live-day commissioning run with a 118~kg of fiducial xenon mass, the LUX experiment has achieved the most sensitive spin-independent WIMP exclusion limits over a wide range of masses. Under the assumption of isospin invariance, this result excludes the low-mass WIMP region where hints of signal have been published. LUX will commence a 300~day run in 2014 that will further improve the WIMP sensitivity by a factor of 5. The increased sensitivity factor is greater than the ratio of exposures due to lower radioactivity backgrounds from the decay of the cosmogenically activated $^{127}$Xe. The sensitivity interpretation at low masses will also benefit from using the novel low-energy calibration of electron and photon yields from the DD neutron source.

\section*{Acknowledgments}
\scriptsize{}
This work was partially supported by the U.S. Department of Energy (DOE) under award numbers DE-FG02-08ER41549, DE-FG02-91ER40688, DE-FG02-95ER40917, DE-FG02-91ER40674, DE-NA0000979, DE-FG02-11ER41738, DE-SC0006605, DE-AC02-05CH11231, DE-AC52-07NA27344, and DE-FG01-91ER40618; the U.S. National Science Foundation under award numbers PHYS-0750671, PHY-0801536, PHY-1004661, PHY-1102470, PHY-1003660, PHY-1312561, PHY-1347449; the Research Corporation grant RA0350; the Center for Ultra-low Background Experiments in the Dakotas (CUBED); and the South Dakota School of Mines and Technology (SDSMT). LIP-Coimbra acknowledges funding from Funda\c{c}\~{a}o para a Ci\^{e}ncia e Tecnologia (FCT) through the project-grant CERN/FP/123610/2011. Imperial College and Brown University thank the UK Royal Society for travel funds under the International Exchange Scheme (IE120804). The UK groups acknowledge institutional support from Imperial College London, University College London and Edinburgh University, and from the Science \& Technology Facilities Council for PhD studentship ST/K502042/1 (AB). The University of Edinburgh is a charitable body, registered in Scotland, with registration number SC005336. This research was conducted using computational resources and services at the Center for Computation and Visualization, Brown University.

We acknowledge the work of the following engineers who played important roles during the design, construction, commissioning, and operation phases of LUX: S. Dardin from Berkeley, B. Holbrook, R. Gerhard, and J. Thomson from UC Davis, and G. Mok, J. Bauer, and D. Carr from Livermore.
We gratefully acknowledge the logistical and technical support and the access to laboratory infrastructure provided to us by the Sanford Underground Research Facility (SURF) and its personnel at Lead, South Dakota. SURF was developed by the South Dakota Science and Technology authority, with an important philanthropic donation from T. Denny Sanford, and is operated by Lawrence Berkeley National Laboratory for the Department of Energy, Office of High Energy Physics.

\normalsize{}

\end{document}